\documentclass[12pt]{iopart}

\usepackage{epsfig}
\input{epsf}

\begin{document}

\title[Inverse Symmetry Breaking in Multi-Scalar Field Theories]
{Inverse Symmetry Breaking in Multi-Scalar Field Theories \footnote {Talk given by Marcus Benghi Pinto at QFEXT05, Barcelona, Spain, Sep. 5-9, 2005.}}

\author{Marcus Benghi Pinto$^1$ and Rudnei O. Ramos$^2$}

\address{$^1$ Departamento de
F\'{\i}sica, Universidade Federal de Santa Catarina, 88040-900
Florian\'{o}polis, SC, Brazil}

\address{$^2$ Departamento de F\'{\i}sica
Te\'orica, Universidade do Estado do Rio de Janeiro, 20550-013 Rio
de Janeiro, RJ, Brazil}
\eads{\mailto{marcus@fsc.ufsc.br}, \mailto{rudnei@uerj.br}}

\begin{abstract}
We review how  the phenomena of inverse
symmetry breaking (and symmetry nonrestoration) may arise in the context of  relativistic as well as nonrelativistic multi-scalar field theories.
We discuss how the consideration of thermal effects on the couplings produce
different transition patterns for both  theories.
For the relativistic case, these effects allow the appearance of inverse
symmetry breaking (and symmetry nonrestoration) at arbitrarily large temperatures.
 On the other hand, the same
phenomena are suppressed in the nonrelativistic case, which is relevant for condensed matter physics. In this case,  symmetry nonrestoration does not happen while inverse symmetry is allowed only to be followed by symmetry restoration characterizing a  reentrant phase. The aim of this paper is to give more insight concerning the, qualitatively correct, results obtained by using one loop perturbation theory in the evaluation of thermal masses and couplings.

\end{abstract}

\submitto{\JPA}

\section{Introduction}

Inverse symmetry breaking (ISB) is the name given to the phenomenon where
a given symmetry may get broken  at high
temperatures. The possibility of such phenomenon taking place in the context
of quantum field theory at finite temperature was first noticed by
Weinberg  \cite{weinberg} who also envisaged that a symmetry which is broken at zero temperature may
not get restored at all at higher temperatures,  a phenomenon called  symmetry
nonrestoration (SNR).

The possibility that some system may acquire lower symmetries as
the temperature increases seems counter intuitive at first sight.
However, there are plenty of real physical systems which do exhibit
phenomena analogous to ISB/SNR. Some examples are given by  the Rochelle salt,
  liquid
crystals, spin glass materials,  compounds
known as the manganites (e.g., $(Pr,Ca,Sr)MnO_3$) and many other systems and
materials, as  recently reviewed in \cite{inverse}.

The idea of ISB/SNR has also called the attention of high energy physicists due to  possibility of their implementation in realistic particle physics models, especially
in the context of high temperature phase
transitions in the early Universe \cite {MR1,prd}. With this purpose, ISB/SNR have been used in applications covering
problems which involve CP violation and baryogenesis, topological defect
formation, inflation, etc
\cite{appl}.

The analysis of   cosmological issues using condensed matter systems is a subject whose importance is growing lately as shown by research programmes such as the COSLAB (Cosmology in the Laboratory) \cite {ray}.
With this purpose we have recently analyzed how ISB/SNR manifest themselves in nonrelativistic theories which may be used in condensed matter physics \cite {prd}. Our aim was to compare the finite temperature symmetry patterns of nonrelativistic models with the ones provided by previous studies concerning the relativistic case.
As we shall review here, these patterns turn out to be completely different when the important thermal effects on the couplings are considered. We will  review, in the next section, ISB/SNR issues concerning  the relativistic model. Then, in section 3,  we will examine the
nonrelativistic case from a perturbatively point of view only. This type of
 analysis has not been performed in detail in Ref. \cite {prd} and will be included here
  since it gives new insights concerning the eventual breakdown
  of perturbation theory as well as  the overall qualitative behavior of the symmetry patterns. Our conclusions are presented in section 4.

  In the companion paper \cite {nois}, we review the nonperturbative results
  for the general nonrelativistic model and present new results concerning an
  explicit application to dilute homogeneous binary Bose gases model
  ($U(1)\times U(1)$ BEC).

\section{Reviewing the original relativistic model}

The theory analyzed by Weinberg \cite{weinberg} consists of a
$O(N_{\phi})\times O(N_{\psi})$ 
invariant relativistic model with two distinct
types of scalar fields, $\phi$ and $\psi$, of $N_\phi$ and
$N_\psi$ components, respectively. The interactions are mediated by a quadratic cross-coupling $\lambda$ as well as by
self-interactions, $\lambda_\phi$ and $\lambda_\psi$.  The Lagrangian density is given by

\begin{equation}
{\cal L}=\frac{1}{2} (\partial_{\mu} \phi)^2  -
\frac {m_{\phi}^2}{2} \phi^2 -
\frac {\lambda_{\phi}}{4!}(\phi^2)^2 +
\frac{1}{2} (\partial_{\mu} \psi)^2  - \frac {m_\psi^2}{2} \psi^2 -
\frac {\lambda_{\psi}}{4!}(\psi^2)^2
-\frac{\lambda}{4} \phi^2\psi^2 \;,
\label{relaction}
\end{equation}

\noindent
where $\phi^2= \sum_i^{N_\phi} \phi_i \phi_i$ and 
$\psi^2= \sum_i^{N_\psi} \psi_i \psi_i$.
Overall boundness  requires the
couplings to satisfy $\lambda_{\phi} >0,\; \lambda_{\psi} > 0$ and
$\lambda_{\phi} \lambda_{\psi} > 9 \lambda^2$. In the one-loop
approximation we can readily compute the thermal masses
for $\phi$ and $\psi$ with the results (at leading order in the high
temperature expansion, $m_\phi/T, m_\psi/T \ll 1$) \cite{weinberg,MR1}

\begin{equation}
M_{i}^2(T) \simeq m^2_i + \frac{T^2}{12} \left [ \lambda_{i}
\frac{1}{2}\left ( \frac {N_{i}+2}{3} \right ) + \lambda
\frac{N_{j}}{2} \right ]\,\,\,,
\;\;\;\;{\rm where} \;\;\;\;{i,j=\phi,\psi}\;\;.
\label{mphi}
\end{equation}

\noindent
{}This equation  shows how ISB/SNR can emerge for
$\lambda < 0$. For example, if one
sets $m_\phi^2$ and $m_\psi^2$ positive to have a symmetric theory at $T=0$,
ISB takes place by choosing

\begin{equation}
|\lambda| > \frac{\lambda_\phi}{N_\psi}\left (\frac{N_\phi+2}{3} \right ) \,,
\label {isbcondition}
\end{equation}
which makes the $T^2$ coefficient of $M_\phi^2 (T)$ negative while the same 
coefficient for $M_\psi^2(T)$ is kept positive, due to the boundness condition.
In this case, high temperatures will induce the breaking 
$O(N_\phi)\times O(N_\psi) \rightarrow O(N_\phi-1)\times O(N_\psi)$ 
at the critical temperature

\begin{equation}
\frac {T_c^2}{m_\phi^2} = 24 \left [ |\lambda | N_\psi - 
\lambda_\phi \left (\frac {N_\phi+2}{3} \right ) \right ]^{-1}\;.
\label{Tcphi}
\end{equation}

On the other hand,  when $m_\phi^2$ and $m_\psi^2$ are initially negative, $\lambda < 0$ and Eq. (\ref {isbcondition} ) holds we have the emergence of SNR in the $\phi$ sector.
At this point one could ask if ISB/SNR are not just artifacts of the naive one loop approximation. This is a rather important point since, in principle, higher orders could bring thermal effects to the couplings in such a way so as to suppress the phenomena.
Let us consider the leading contribution, to the couplings,
in the high temperature approximation

\begin{equation}
\lambda_{i}(T) \simeq \lambda_{i} + \frac {3}{8 \pi^2}
\ln \left ( \frac {T}{M_0} \right )\left[  \frac {1}{2}
\left ( \frac {N_{i}+8}{9} \right ) \lambda_{i}^2 +
\frac {N_{j}}{2} \lambda^2 \right ]\;,
\label{lphiT}
\end{equation}

and

\begin{equation}
\lambda(T) \simeq \lambda  + \frac {\lambda}{8 \pi^2}
\ln \left ( \frac {T}{M_0} \right ) \left [  \frac {1}{2}
\left ( \frac {N_{\phi}+2}{3} \right )  \lambda_{\phi} +
\frac {1}{2} \left ( \frac {N_{\psi}+2}{3} \right )
\lambda_{\psi} + 2 \lambda\right ]  \;,
\label{lphipsiT}
\end{equation}

\noindent
where $M_0$ is a regularization mass scale.
Then, the thermal masses can be rewritten as $M_i^2(T) = m_i^2+ (T^2/12) \Delta_i(T)$ where $\Delta_i(T) =  \lambda_i (T) (N_i+2)/6 +  \lambda (T) N_j/2$.

\noindent
It turns out that  these improved results for $M_i^2(T)$ also allow for the appearance of ISB/SNR \cite {MR1, roos}. However, at this point one could raise an objection following the fact that those results are still perturbative and, as well known, perturbation theory eventually breaks down for field theories at finite temperatures. However, nonperturbative evaluations carried out with the Wilson Renormalization Group equations  \cite {roos} as well as with the linear $\delta$ expansion \cite {MR1}  show the correctness of those results in the qualitative sense. Refs. \cite {eles} also support the
occurrence of those exotic phenomena.
That is, the inclusion of thermal effects on the couplings does not exclude the
possibility of SNR/ISB occurring, at high temperatures,  in
$O(N_\phi) \times O(N_\psi)$ scalar  relativistic models.

\section{The nonrelativistic model and the appearance of reentrant phases}
\label{NR}
We now turn our attention to the analysis of  SNR/ISB phenomena
 in the  nonrelativistic limit. Let
us first recall some fundamental differences between relativistic and
nonrelativistic theories that will be important in our analysis.
{}Firstly, the obvious reduction from Lorentz to Galilean invariance.
Secondly, it should be noted that in the nonrelativistic description
particle number is conserved and so, only complex fields are allowed.
This second point will be particularly important to us since, for the
processes entering  the evaluation of the effective couplings  only those that do not change
particle number (the elastic processes) will be allowed. Another important difference between
relativistic and nonrelativistic models concerns the structure of the
respective propagators. While the relativistic propagator allows for
both forward and backward particle propagation (which is associated to
particles and anti-particles, respectively), the nonrelativistic
propagator of scalar theories at $T=0$ only allows for  forward propagation.  Note however that the
structure of the propagators  in a
thermal bath includes both backward and forward propagation
\cite{mahan}. The nonrelativistic lagrangian density is given by \cite {prd}
\begin{eqnarray}
{\cal L} &=&
\Phi^* \left(i \partial_t + \frac{1}{2m_\Phi}\nabla^2
\right) \Phi - \kappa_\Phi \Phi^* \Phi
- \frac{g_\Phi}{3!} (\Phi^* \Phi)^2
\nonumber \\
&+& \Psi^* \left(i \partial_t + \frac{1}{2m_\Psi}\nabla^2
\right) \Psi - \kappa_\Psi \Psi^* \Psi
- \frac{g_\Psi}{3!} (\Psi^* \Psi)^2- g (\Phi^* \Phi) (\Psi^* \Psi)\;.
\label{NRL}
\end{eqnarray}

\noindent
 The one body parameters $\kappa_i$, with $i=\Phi,\Psi$,
can account for external potentials, or chemical potentials (in the grand
canonical formalism) which is important for Bose-Einstein condensation (BEC).
The $m_i$  represents the (atomic)
masses. {}For the BEC case, the couplings are related to the $s$-wave 
scattering length.
Here, we will not attach any specific role to any of those parameters, 
leaving this to the companion paper \cite {nois}.
The numerical
factors and signs in the one and two-body potential terms
have been chosen in such a way so that the potential is
analogous the one considered in (\ref{relaction}). The
boundness condition for the model (\ref{NRL}) is analogous
to that of the relativistic model
(\ref{relaction}), 
requiring $g_\Psi > 0$, $g_\Phi >0$ and
$g_\Psi g_\Phi > 9 g^2$.
In addition, notice that for the nonrelativistic limit to be valid, 
one must keep $T \ll m_i$. In general,
for nonrelativistic systems,  the masses $m_i$ are of order
of typical atomic masses, $m_i \sim {\cal O}(1-100) {\rm GeV}$. 
At the same time, the
typical temperatures in condensed matter systems are at most of order of
a few eV. Therefore, this condition will always hold for the ranges of 
temperature we will be interested in below.

{}For multi-component fields, (\ref{NRL}) is a nonrelativistic
multi-scalar model with symmetry $U(N_\Phi) \times U(N_\Psi)$ that is
the analogue of the original relativistic model (\ref{relaction}).
{}For simplicity, in the following we set $N_\Phi=N_\Psi=1$, corresponding 
to an $U(1)\times
U(1)$ symmetric model. In this case, one can  write the complex fields in
terms of real components as $\Phi = (1/\sqrt{2}) ( \phi_1 + i \phi_2)$ and
$\Psi = (1/\sqrt{2}) ( \psi_1 + i \psi_2)$.

\noindent
The nonrelativistic model, Eq. (\ref{NRL}), falls in the same class of
universality as the $O(2) \times O(2)$ relativistic case.
At the one loop level, one can then write $\kappa_i$  in the high temperature
approximation  as

\begin{equation}
\kappa_i(T) \simeq   \kappa_i +
\left ( \frac {T}{2 \pi } \right )^{3/2} \zeta (3/2) \Delta^{\rm NR}_{i} \;,
\label{kidel}
\end{equation}
defining the quantity $\Delta^{\rm NR}_{i}=  (2/3) g_i m_i^{3/2}  +  g m_j^{3/2}$.

\noindent
Then, the  critical temperature for symmetry
restoration/breaking is

\begin{equation}
T_{c,i}^{\rm NR} = 2\pi
\left [ \frac {-\kappa_i}{\Delta^{\rm NR}_i\zeta(3/2)} \right ]^{2/3}\;.
\label{tc}
\end{equation}
As in the relavistic case, the exotic transition patterns arise for $g<0$. For example,
$\Delta^{\rm NR}$ signals the appearance of ISB or SNR in the $\Psi$ sector if one chooses $|g| > (2/3) g_\Psi$. We have seen that in the relativistic case this possibility survives to the inclusion of thermal effects on the couplings. However, as already discussed, the nonrelativistic contributions which are allowed in the computation of the four point Green's functions may produce $g_i(T)$ and $g(T)$ which differ drastically from the relativistic $\lambda_i(T)$ and $\lambda(T)$. In fact, the one loop evaluation, for high-$T$, gives \cite {prd}

\begin{equation}
g_{i}(T) \simeq g_{i} - \frac{m T}{12 \pi} \sqrt{ \frac{2 m}{\kappa}}
 \left( 5 g_i^2 + 9 g^2 \right) + {\cal O}(\kappa/T)\;,
\label{gPhiT}
\end{equation}

and

\begin{equation}
g(T) \simeq  g -  \frac{m T}{4 \pi} \sqrt{ \frac{2 m}{\kappa}}
 g \left(g+ \frac{2 g_\Phi}{3} +  \frac{2 g_\Psi}{3}
\right) + {\cal O}(\kappa/T) \;,
\label{gPhiPsiT}
\end{equation}

\noindent
where we have set $m_\Phi=m_\Psi=m$ and $\kappa_\Phi=\kappa_\Psi=\kappa$.
Note from the equations above that
the effective couplings in the nonrelativistic theory have a much
stronger dependence with the temperature than those in the equivalent
relativistic theory. We therefore expect to see larger deviations at high
temperatures for the effective couplings as compared with the same case
in the relativistic problem (by high temperature we mean $k_i(0)<<T<<m_i$). 
It is
also evident from the analysis of higher loop corrections to the
effective couplings in the nonrelativistic model that all bubble like
corrections contribute with the same power in temperature as the
one-loop terms, which can easily be checked by simple power-counting in
the momentum. A side effect of this is the breakdown, at high
temperatures, of the simple one-loop perturbation theory applied here.
Another symptom  is the apparent running of the effective
self-couplings, shown above, to negative values for sufficiently high
temperatures given by $T^{\rm neg} \sim {\rm min} \left( 12
\pi \sqrt{\kappa/(2 m^{3})} g_\Phi/(5 g_\Phi^2 + 9g^2),\; 12 \pi
\sqrt{\kappa/(2 m^{3})} g_\Psi/(5 g_\Psi^2 + 9g^2) \right)$. At the same time,
perturbation theory breaks down when the temperature dependent boundness condition,
$R^{\rm NR} (T)= g_\Psi(T) g_\Phi(T)/[3g(T)]^2$, becomes smaller than the unity.
Nevertheless, it is easy to check that (for the parameters adopted
below) the results obtained by just plugging 
Eqs. (\ref {gPhiT})-(\ref {gPhiPsiT}) above into (\ref {kidel}) already
shows a drastic qualitative difference between this simple improved
approximation and the naive perturbative evaluation given  by Eq.
(\ref {kidel}).  This can be seen from figures 1 and 2 where 
$\kappa_\Phi(T)$ and
$\kappa_\Psi(T)$ are displayed using temperature dependent 
and temperature independent couplings.
\begin{figure}[htb]
\vspace{0.5cm}
\centerline{\epsfig{figure=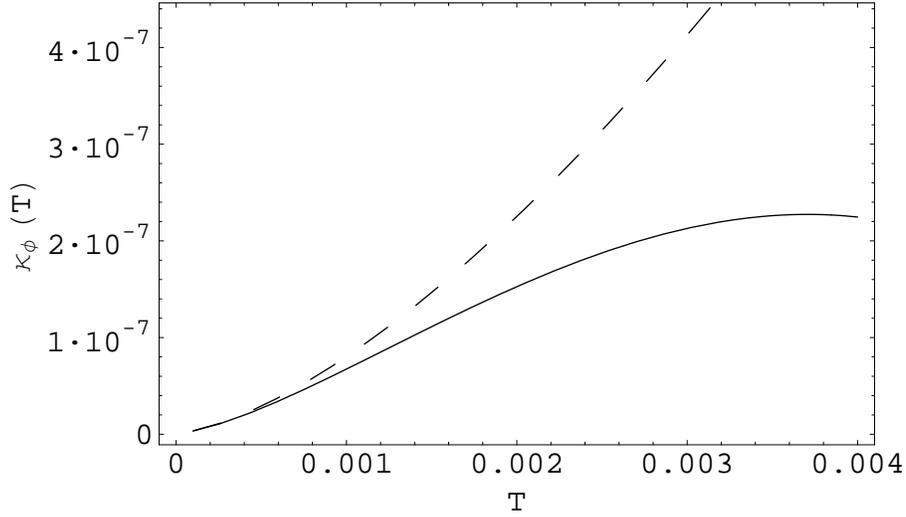,angle=0,width=12cm}}
\caption[]{\label{fig1} The quantity $\kappa_\Phi(T)$ obtained by using
bare couplings (dashed line) as well as temperature dependent 
couplings (continuous line). Both, $\kappa_\Phi(T)$ and $T$ are 
given  in  eV unities. }
\end{figure}
\begin{figure}[htb]
\vspace{0.5cm}
\centerline{\epsfig{figure=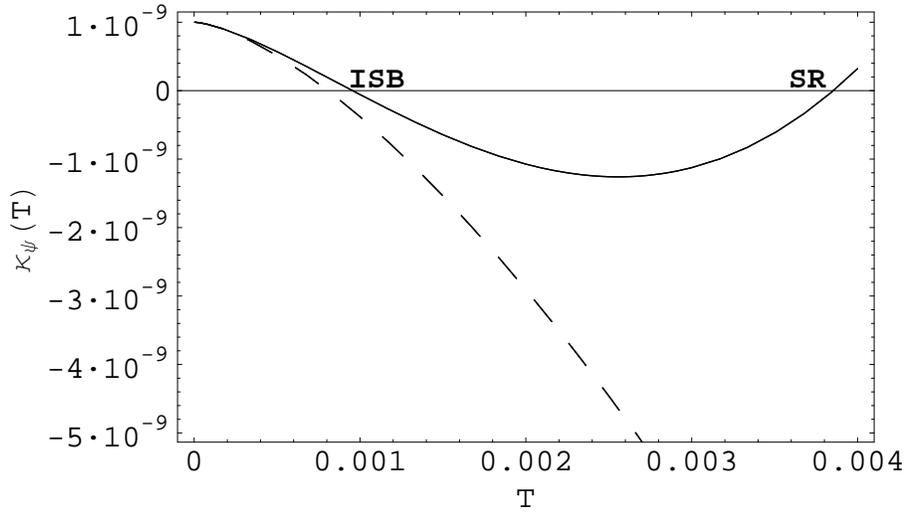,angle=0,width=12cm}}
\caption[]{\label{fig2} The quantity $\kappa_\Psi(T)$ obtained by using
bare couplings (dashed line) as well as temperature dependent couplings (continuous
line). In the second case one observes ISB followed by SR characterizing a reentrant phase. Both, $\kappa_\Psi(T)$ and $T$ are given in eV unities.}
\end{figure}

The parameter values are $g_\Phi(0)= 8 \times 10^{-16} \,\,{\rm eV}^{-2}$, $g_\Psi(0)= 7 \times 10^{-17} \,\,{\rm eV}^{-2}$, $g(0)= -5.5 \times 10^{-17} \,\,{\rm eV}^{-2}$, $m_\Phi=m_\Psi= 1\,\, {\rm GeV}$ and $\kappa_\Psi(0)=\kappa_\Phi(0)
= 1 \,{\rm neV}$. Using these values and Eqs. (\ref {gPhiT})-(\ref {gPhiPsiT}) in the evaluation of $\Delta^{NR}$ one sees the appearance of a reentrant phase on the $\phi$  sector characterized by ISB (at $T_c^{\rm ISB} \simeq 1 \times 10^{-3} \, {\rm eV}$) followed by SR (at $T_c^{\rm SR} \simeq 4 \times 10^{-3} \, {\rm eV}$).
To assess the validity of perturbation theory one must be sure that those values of $T_c^{\rm ISB}$ and $T_c^{\rm SR}$ are smaller than the temperature which signals that the potential becomes unbounded via $g_\Psi(T) < 0$, $g_\Phi(T) <0$ or $R^{\rm NR}(T) <1$.
For the parameter values considered here, $g_\Psi(T)$ reaches negatives values, before $g_\Phi(T)$, at
$T^{\rm neg} \simeq 6.6 \times 10^{-3}$ eV. At the same time, figure 3 shows that $R^{\rm NR} (T) \le 1$ at a smaller temperature, namely $T^{\rm unbound} \simeq 4.675 \times 10^{-3} \, {\rm eV}$.
\begin{figure}[htb]
\vspace{0.5cm}
\centerline{\epsfig{figure=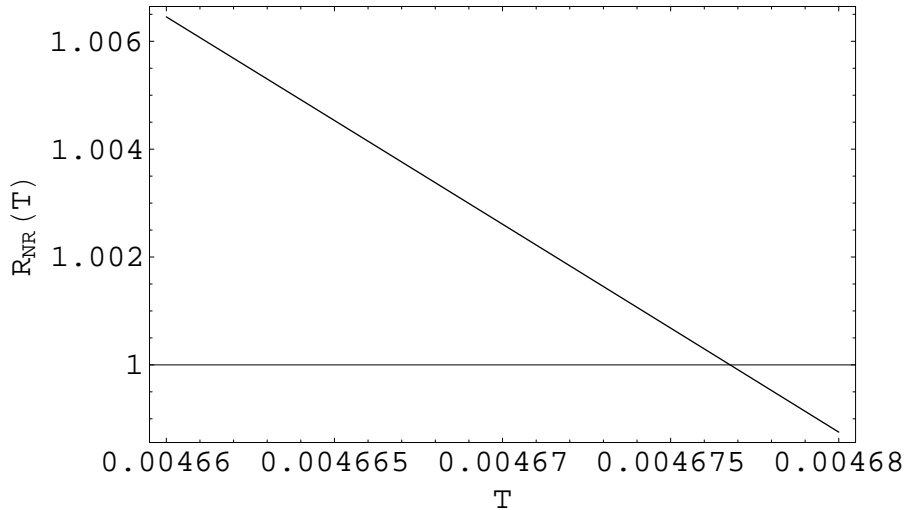,angle=0,width=12cm}}
\caption[]{\label{fig3} The dimensionless quantity $R^{\rm NR}(T)= g_\Psi(T) g_\Phi(T)/[3g(T)]^2$
 as a function of T(eV). One sees that $R^{\rm NR}(T) \le 1$ at $T^{\rm unbound} \simeq 4.675 \times 10^{-3} \, {\rm eV}$.}
\end{figure}
Therefore, the appearance of the important reentrant phase, 
in the $\psi$ sector, happens at critical temperatures smaller than the 
ones which cause the instability of the potential.
So, at least qualitatively, the strictly perturbative result presented 
in this paper is already rather satisfactory and has been  confirmed by
a nonperturbative resummation \cite {prd}. 

The results presented here
show that the phenomenon of reentrant phases, like those of ISB/SNR, observed in the relativistic
models are not exclusively nonperturbative phenomena but also feasible
within perturbation theory.
{}Furthermore, reentrant phases in nonrelativistic models like (\ref{NRL})
can be seen as a consequence of the interplay of the different
energy scales available, that can be expressed in terms of
 $\kappa_i$, $m_i$ and the (dimensionfull) couplings,
$g$ and $g_i$. At the same time, the relativistic model
characterized  only by the scales $m_i$, symmetry breaking/restoration
phenomena is only accessible at high energy scales, $T\gg m_i$, as
clearly shown by the general critical temperature result,
Eq. (\ref{Tcphi}), for perturbative values of coupling constants.
The fact that the emergence of reentrant phases is a genuine feature of  nonrelativistic
models which is not affected by relativistic corrections can be seen, for instance, by considering the expansion of the relativistic dispersion
relation

\begin{equation}
\omega_i = (k^2 + m_i^2)^{1/2} \sim m + \frac{k^2}{2 m_i}
-\frac{1}{2m_i} \left(\frac{k^2}{2 m_i}\right)^2 + \ldots
\label{dispersion}
\end{equation}

\noindent
Since the temperature roughly gives the magnitude of the kinetic
energy, the third term in (\ref{dispersion}) relative to the 
second one is of order ${\cal O}(T/m_i) \ll 1$. Higher order corrections
to the nonrelativistic term are, therefore, negligible for
the results obtained, e.g., for the  reentrant phase temperatures,
$T^{\rm ISB}$ and $T^{\rm SR}$, shown for instance in {}Fig. \ref{fig2}.

\section{Conclusions}
\label{conclusions}

We have seen how symmetry nonrestoration and inverse symmetry
breaking may take place, at arbitrarily large temperatures, in
multi-field scalar relativistic and nonrelativistic theories. These
counter intuitive phenomena appear due to the fact that the crossed
interaction can be negative while the models are still bounded from
below. We have reviewed that, in the relativistic case, ISB/SNR survive
the inclusion of thermal effects on the couplings. This qualitative result,
first obtained perturbatively is confirmed by nonperturbative evaluations 
\cite {MR1, prd, roos, eles}.
Then, we have analyzed   the possibility
of obtaining such transition patterns in the nonrelativistic case. 
The naive use
of temperature independent couplings allows ISB (and SNR) but the 
inclusion of the leading
thermal contribution already causes  of  a drastic difference. Namely,
 ISB can show up in the nonrelativistic case  only via the 
appearance of reentrant phases, as indeed observed in many real 
condensed matter systems. In the present work, our aim was to 
perform a deeper investigation of the perturbative results than 
the one which was done in Ref. \cite {prd}. We have shown that 
perturbation theory is capable of predicting the right transition 
behavior provided one stays within its limit of applicability. 
Indeed,  when thermal corrections are included, the qualitative
results obtained  here  are verified
by nonperturbative calculations as we review in Ref. \cite{nois} 
where the  Bose-Einstein condensation problem for binary 
gases is considered.

\ack

The authors were partially supported by Conselho Nacional de
Desenvolvimento Cient\'{\i}fico e Tecnol\'{o}gico (CNPq-Brazil).
ROR was also partially supported by FAPERJ.
We also would like to thank the organizers of the QFEXT05 workshop in
Barcelona.


\section*{References}

\begin{thebibliography}{99}


\bibitem{weinberg} Weinberg S 1974 {\it  Phys. Rev.} D {\bf 9} 3357
\bibitem{inverse} Schupper N and Shnerb N M  {\it Preprint} cond-mat/0502033

\bibitem{MR1}Pinto M B and Ramos R O 2000 {\it Phys. Rev.} D {\bf 61} 125016

\bibitem{prd}Pinto M B , Ramos R O  and Parreira J E  2005 {\it Phys. Rev.} D {\bf 71} 123519

\bibitem{appl}Mohapatra R N  and Senjanovic  G 1979 {\it Phys. Rev. Lett.}
{\bf 42} 1651;
Dodelson S  and  Widrow L M 1990 {\it Phys. Rev. Lett.} {\bf 64},
340; Bajc B and  Senjanovi\'c G 1997 {\it Nucl. Phys. Proc.
Suppl.} A {\bf 52} 246

\bibitem{ray} Rivers R J, {\it Preprint}  cond-mat/0412404,
published in: {\it Proceedings of the National Workshop on Cosmological
Phase Transitions and Topological Defects} (Porto, Portugal, 2003), ed.
Girard T A (Grafitese, Edificio Ciencia, 2004), 11-23.



\bibitem{nois} Ramos R O and Pinto M B, {\it Symmetry Aspects in
    Nonrelativistic Multi-Scalar Field Models and Application to a Coupled
    Two-Species Dilute Bose Gas}, in press J. Phys. A (2006).

\bibitem{roos} Roos T 1996 {\it Phys. Rev.} D {\bf 54} 2944

\bibitem{eles}Jansen K and Laine M 1998 {\it Phys. Lett.} B {\bf 435} 166;
Bimonte G, I\~niguez D, Taranc\'on A and  Ullod C L 1993
{\it Nucl.Phys.} B{ \bf 559} 103


\bibitem{mahan}Mahan G D {\it Many-particle Physics}
(Plenum, New York, 1981).

\end{thebibliography}
\end{document}